\documentclass[acmsmall]{acmart}

\usepackage{subcaption}
\usepackage[utf8x]{inputenc}
\usepackage{pifont}
\usepackage{multirow}

\usepackage[normalem]{ulem}
\useunder{\uline}{\ul}{}

\newcommand{\citelong}[1]{\citeauthor{#1} (\citeyear{#1})~\cite{#1}}

\setcopyright{acmlicensed}
\copyrightyear{2024}
\acmYear{2024}
\acmDOI{XXXXXXX.XXXXXXX}

\acmJournal{TOMM}

\begin{document}

\title{Task Presentation and Human Perception in Interactive Video Retrieval}

\author{Nina Willis}
\email{ninamari.willis@uzh.ch}
\orcid{0009-0000-2491-4948}
\affiliation{%
  \institution{University of Zurich}
  \city{Zurich}
  \country{Switzerland}
}

\author{Abraham Bernstein}
\email{bernstein@ifi.uzh.ch}
\orcid{0000-0002-0128-4602}
\affiliation{%
  \institution{University of Zurich}
  \city{Zurich}
  \country{Switzerland}
}

\author{Luca Rossetto}
\email{rossetto@ifi.uzh.ch}
\orcid{0000-0002-5389-9465}
\affiliation{%
  \institution{University of Zurich}
  \city{Zurich}
  \country{Switzerland}
}

\begin{abstract}
Interactive video retrieval is a cooperative process between humans and retrieval systems.
Large-scale evaluation campaigns, however, often overlook human factors, such as the effects of perception, attention, and memory, when assessing media retrieval systems.
Consequently, their setups fall short of emulating realistic retrieval scenarios.
In this paper, we design novel task presentation modes based on concepts in media memorability, implement the pipelines necessary for processing target video segments, and build a custom experimental platform for the final evaluation.
In order to study the effects of different task representation schemes, we conduct a large crowdsourced experiment.
Our findings demonstrate that the way in which the target of a video retrieval task is presented has a substantial influence on the difficulty of the retrieval task and that individuals can successfully retrieve a target video segment despite reducing or even altering the provided hints, opening up a discussion around future evaluation protocols in the domain of interactive media retrieval.
\end{abstract}




\maketitle

\section{Introduction}

In an era characterized by an ever-expanding wealth of information online, there is an increasing need for efficient methods of browsing and retrieving specific content.
While retrieval systems have made impressive advancements, we still see that, more often than not, the retrieval process is an interactive one \cite{rossetto2020interactive}.
Retrieval systems present a set of results based on search queries, but these results may or may not provide what the searcher is looking for, whether it is due to limitations within the system or the quality of the queries themselves.
In many cases, searchers scan the results and refine their queries in an iterative process, essentially engaging in a feedback loop with the retrieval system until they discover what they seek. 

In order to enhance such interactive retrieval systems for optimal content retrieval, it is imperative to evaluate their performance and identify effective techniques and technologies.
An inherent challenge in these evaluations is accounting for the human factor in the retrieval process and simulating realistic interactive retrieval scenarios. One such scenario is the \textit{known-item search (KIS)} task, where the searcher aims to retrieve something they have encountered before from a database where they know the item exists.
Video search evaluations, such as the Video Browser Showdown (VBS), attempt to approximate this scenario by repeatedly presenting the target segment and asking participants to find the video as quickly as possible using their retrieval system \cite{schoeffmann2019video}.
This, however, is not a realistic representation of the situation.
In real life, certain details usually go unnoticed or are forgotten by the time people want to retrieve the video again. 
Repetitive exposure to the target video in a time-sensitive search environment may encourage heightened attention to elements typically unnoticed or unremembered, such as a watermark in the corner of the video or minuscule details in the landscape.

Although media retrieval evaluations are designed to assess state-of-the-art systems for their efficiency and usability, none so far have considered the impact of human perception, attention, and memory on the interactive retrieval process.
The central question then revolves around how to represent search tasks to replicate a KIS scenario for authentic video retrieval evaluations.

In this paper, we design and construct various preprocessing pipelines to create novel task representations that aim to encompass the intricacies of perception and memory effects.
We then explore the implications of employing these representations in KIS evaluations through a large-scale crowdsourced user experiment conducted on a custom platform.
We find that varying the representation of the target of a retrieval task has a clear effect on retrieval performance.
Specifically, filtered target images that simulate effects similar to those caused by attention and memory show a slight decrease in retrieval performance.
They do, however, perform better than purely textual descriptions, suggesting that processing even partial visual information is easier than imagining it based on a text.
A synthetic, generated target seems even to aggravate the performance, possibly emulating misremembering a target. 
These findings shed new light on the effects of the design of retrieval benchmarks and the real-world scenarios they are trying to model.
Future iterations of relevant benchmarks might be adjusted in order to reflect their target scenarios more accurately.

The remainder of this paper is structured as follows: Section~\ref{sec:related} discusses related work around multimedia retrieval evaluations and memorability estimation. Section~\ref{sec:pipelines} introduces our video processing pipelines that aim to simulate certain perception and memory effects. These pipelines are then tested in a preliminary evaluation in Section~\ref{sec:preliminary}. Section~\ref{sec:setup} outlines the setup of our large crowdsourced experiment, the results of which are presented in Section~\ref{sec:results}. Section~\ref{sec:discussion} discusses the insights that we gain from the experiment before we offer some concluding remarks in Section~\ref{sec:conclusion}.

\section{Related Work}
\label{sec:related}
This section delves into works related to media retrieval evaluations, focusing on interactive video search, followed by a review of research on human attention and memory, focusing on media memorability.

\subsection{Media Retrieval Evaluations}
The evaluation of information retrieval techniques came about from a need for a controlled comparison of the proposed methods and, essentially, to consolidate findings that researchers can use to advance the field further.
Spreading from textual search evaluations \cite{harman1993overview}, evaluation campaigns have been established for image retrieval \cite{clough2004clef}, as well as for video content search \cite{smeaton2001trec}.
The current space of content-based video search evaluations consists mainly of known-item search (KIS) tasks and ad-hoc video search (AVS) tasks \cite{lokovc2022task}.
Originating from library science, KIS tasks simulate the situation in which the searcher knows of a target scene that exists in the collection and wishes to retrieve it again.
This task type was introduced to the video search domain in the Text Retrieval Conference (TREC) Video Track 2001 \cite{smeaton2001trec}, later known as TREC Video Retrieval Evaluation (TRECVID) \cite{awad2020trecvid}.
TRECVID also introduced the AVS task, a subject-wide search in which the searcher aims to find as many scenes that belong to a given topic as possible.
Early media retrieval evaluations had been focused primarily on automatically assessing the effectiveness of the search systems.
In a realistic setting, media retrieval tends to be an interactive process between the human searcher and the retrieval system.
In order to evaluate the interaction and interface aspects of video retrieval, VideOlympics \cite{snoek2008videolympics} was established in 2007 as a live competitive event with AVS tasks.
Expanding on this sentiment, Video Browser Showdown (VBS) was introduced in 2012 as a live interactive video retrieval evaluation event \cite{schoeffmann2019video}.
The annual competition evaluates both KIS and AVS tasks, with KIS tasks being further divided into visual KIS, where the actual target scene is presented in a repetitive manner, and textual KIS, where only a textual description of the target scene is presented to the searchers.
Search performance and usability are assessed through the participation of both expert and novice searchers.

The VBS has evolved over the years.
The dataset size, for instance, has grown significantly since 2012 \cite{schoeffmann2019video}.
To better analyze and learn from the results of the competition, VBS introduced interaction logging in 2019 \cite{rossetto2020interactive}, followed by additional result set logging in 2020 \cite{lokovc2021reign}.
These logs facilitate understanding the effectiveness of various query methods and ranking models.
In 2021, VBS held its first fully remote competition~\cite{heller2022interactive} by adopting the Distributed Retrieval Evaluation Server (DRES)~\cite{rossetto2021system}.
Designed to facilitate the evaluation of interactive multimedia retrieval systems in both on-site and distributed settings, DRES supports all aspects of the evaluation, including task configuration and logging.
It was later extended to enable asynchronous evaluations, relaxing both the locality and time constraints \cite{sauter2022asynchronous}.
The remote setting of VBS 2021 demonstrated the feasibility of conducting a fully virtual evaluation of interactive systems.
While the virtual setting has the obvious benefits of lower cost and no geographical barriers to participation, some aspects of the on-site setting, such as networking and novice sessions, could not be fully replicated \cite{heller2022interactive}, suggesting that rather than fully replacing on-site competitions, distributed evaluations can supplement them to enable more frequent evaluation opportunities or even be used to experiment with new task types \cite{sauter2022asynchronous}.
With the establishment of a flexible server, it is now possible to configure tasks in different ways and to more easily explore the task category space for video search evaluations, much of which is still underexplored \cite{lokovc2022task}. 

The incorrectness or incompleteness of the searcher's understanding of their target has been a long-identified challenge of known-item search \cite{lee2006known}, but little has been done to address this in retrieval system evaluations. The visual KIS task, for instance, does not take into consideration the effects of human perception and memory.
However, presenting multiple scenes before the actual competition requires more time and participants to adjust for individual variations.
A possible approach to address this issue is to model the effects of attention and memory explicitly and to simulate them in the presented content \cite{lokovc2022task}.
An initial step in this direction uses a saliency mask based on eye-gaze information to predict and visually degrade information that would not have been attended to in the target video \cite{rossetto2021considering}.
No substantial difference in retrieval performance was found between using the original unfiltered videos and the filtered ones, indicating that some information can be removed without negatively impacting the solubility of the video retrieval task.
While this prior work simulated the effects of attention, it did not explicitly look into the effects of human memory and the memorability of media content, nor did it simulate the effects of auditory attention or memory on the audio track other than a content-independent filter, which leaves room for further exploration.

\subsection{Media Memorability}
After the initial sensory memory stage, in which perceived information is briefly stored, attention facilitates the advancement of certain information into short-term or working memory and eventually into long-term memory \cite{baddeley1999essentials}.
The short-term memory component has a very limited capacity but high recognition accuracy, while the long-term memory component has lower recognition accuracy but greater capacity and stability \cite{phillips1977components}.
Working memory and attention have a very close relationship with object representation in visual working memory appearing to be overwritten when attention is directed to a new object within the same category \cite{olsson2005visual}.
In contrast, the long-term memory component can store many objects, their overall categorical information, and details of those objects \cite{brady2008visual}.
These item-specific details, however, cannot be stored in long-term memory without the support of preexisting conceptual knowledge.
Visual information capacity in long-term memory, hence, depends more on conceptual structure than perceptual distinctiveness between presented objects \cite{konkle2010conceptual}, emphasizing the importance of semantics in human memory.
Scene representation in memory has also been found to have a high level of fidelity, suggesting that scenes and objects are represented at the same level of abstraction in visual long-term memory \cite{konkle2010scene}.

Memory has been found to decay deterministically, making it possible to predict the effects of memory decay with some forgetting function \cite{gold2005visual}.
The degradation of memory, however, does not occur uniformly.
Certain kinds of information, such as the objects in an image and their relative locations, stay in long-term memory. Others, such as details of objects within a scene and overall spatial composition, are not as well retained \cite{mandler1977long}.
More recently, memory representations of visual scenes have been found to lose not only their high-level details but also their low-level visual qualities, such as color saturation and luminance \cite{cooper2019memories}.

Memorability was found to be an intrinsic property of images, which means that certain types of visual content are more memorable than others, regardless of viewer demographics, context, or other high-level qualities such as aesthetics or interestingness \cite{isola2011makes,isola2013makes}.
This sparked research on media memorability, including image memorability and, more recently, video memorability.
Video memorability data collection began with the collection of brain activity data \cite{han2014learning}, which demonstrated useful findings but is difficult to scale and generalize.
Designed based on the semantic storing of information in human memory, a more scalable protocol was proposed, in which participants were asked textual recall questions about the videos they remember \cite{shekhar2017show}.
Textual recall prompts, however, do not fully represent visual recall and pose several limitations, such as exclusion by language barrier.
To address these issues, another protocol was proposed in which long-term video memorability annotations were collected through participants' prior memory of well-known movies \cite{cohendet2018annotating}, but this, in turn, had its own limitations, such as limited content choice and reliance on participants' subjective judgments.
Collected by measuring memorability within a few minutes and 24-72 hours after memorization, VideoMem \cite{cohendet2019videomem} was introduced as a large-scale dataset composed of videos with both short-term and long-term memorability scores.
Finally, Memento10k was introduced as a dynamic video memorability dataset, enabling the prediction of not only the memorability score of a video but also its rate of decay over time \cite{newman2020multimodal}.
The VideoMem and Memento10k datasets are often used in testing media memorability estimation models, such as in the MediaEval video memorability prediction task \cite{sweeney2022overview}, which continues to challenge researchers today.
In the following, we focus on visual and audio memorability. 

\subsubsection{Visual Memorability}
Ever since memorability was identified as a stable and intrinsic property of an image \cite{isola2011makes}, many approaches have been made towards predicting image memorability, which transfers well into the multimodal realm of video.
Since visual memory is more dominant than auditory memory \cite{cohen2009auditory, bigelow2014achilles}, features related to the visual track are more strongly predictive of overall video memorability than the audio track.
While many different features have been explored within the visual domain, most features and their relationships to memory can perhaps be explained through the lenses of \emph{saliency} and \emph{semantics}.

\paragraph{Saliency}
The role of attention in visual perception has been studied extensively, with the consensus being that humans are guided by a combination of bottom-up pre-attentive processes, which direct perception toward salient stimuli, and top-down processes, which are driven by the viewer's state of mind, including context, task, goals, and memory \cite{healey2011attention}.
Without attention, significant changes to a scene \cite{rensink1997see} or even the presence of visually salient objects can go completely unperceived \cite{simons1999gorillas} and thus not encoded into memory.
Attention and memory, in fact, have an interdependent relationship.
While attention determines what will be encoded, memory guides what should be attended \cite{chun2007interactions}.
Since attention guides what information gets stored in memory, saliency is often explored as a feature for media memorability prediction.
Within the realm of video memorability prediction in particular, \citelong{kar2017makes} and \citelong{shekhar2017show} identified visual saliency early on as a key feature contributing to overall video memorability.
Although not discussed extensively here, salient aspects in video may include features such as motion onset, which can attract attention \cite{abrams2003motion}. 

Guided by theories of human attention, several studies combine bottom-up visual saliency prediction with top-down object-level approaches to estimate attentional mechanisms for the prediction of image memorability.
One method identifies bottom-up saliency map coverage and object contrast level as two attention-related features that can complement and/or replace other low-level features for memorability prediction \cite{mancas2013memorability}.
Another uses spatial histograms based on object-saliency maps of images, generated by replacing each detected object with its average bottom-up visual saliency \cite{wang2015investigation}.
These studies confirm that although it is able to improve predictions of memorability when taken with other features, saliency alone is not a perfect predictor of memorability \cite{celikkale2013visual, mancas2013memorability, fajtl2018amnet}.
While saliency can help predict object memorability in simple scenes, it is not as strong of a predictor when there are many points of interest \cite{dubey2015makes}, suggesting that memorability and salience are affected by different factors.
Visually salient regions may not be the most memorable ones.
Likewise, memorable regions are not necessarily the most salient ones. 

However, the role of attention in memorability should not be overlooked.
Again, objects must be attended to in order to be perceived and remembered, and memorable image regions correlate well with real visual fixations.
Highly memorable images tend to have more consistent human fixations, with only one or a few points of focus \cite{khosla2015understanding, mancas2013memorability, wang2015investigation, lyu2020overt}.
Memorable regions correlate with longer visual fixation durations \cite{mancas2013memorability}.
Unlike bottom-up saliency maps, real gaze fixation maps actually correlate quite well with memorability maps \cite{khanna2019memorability}.
The limited predictive power of saliency so far may be due to the fact that bottom-up saliency maps do not perfectly predict visual attention in real-world scenes and that top-down attentional mechanisms are complex to model accurately \cite{wu2014guidance}.
AMNet \cite{fajtl2018amnet}, an attention-based memorability estimation network, produces three visual attention maps, each conditioned on the one before it, to predict image memorability.
Attention appears to move from pre-attentive visual saliency, generally more influenced by center bias, towards regions more responsible for memorability, as if through top-down attentional mechanisms.
With the growth of available eye-tracking data, deep learning approaches can now be used to model attention and predict video saliency maps based on real gaze information \cite{jiang2018deepvs}, which enables the learning of more complex patterns of human attention.

\paragraph{Semantics}
Top-down attentional processes are guided by semantic information accessed from long-term memory \cite{wu2014guidance}.
However, attention only partially explains the relationship between scene semantics and memorability, suggesting that semantic information relevant to attention and semantic information relevant to memorability are overlapping but different \cite{lyu2020overt}.
Semantics have been found time and time again to play an important role in media memorability, with foregrounds, especially humans and human-scale objects, contributing most positively to memorability, and backgrounds, especially exteriors and natural scenes, contributing most negatively to memorability \cite{isola2011makes, khosla2015understanding}.
Objects in the scene appear to be particularly important, as image memorability is greatly influenced by the memorability of its most memorable object \cite{dubey2015makes}.
High-level object and scene semantics were found to be the most predictive features of overall image memorability \cite{isola2011makes} and can be exploited even without the manual annotation of object labels to predict memorable image regions \cite{khosla2012memorability}.
In line with findings from studies on human memory \cite{konkle2010conceptual}, conceptual categories influence which regions of an image will be remembered.
Some categories of objects are found to be more memorable than others and even affected by the presence of other objects at different rates of decay \cite{dubey2015makes}.

Semantic features are also prevalent in video memorability prediction.
The first video memorability prediction framework, which used functional magnetic resonance imaging (fMRI) to learn memorability from brain activity, found that the occurrence likelihood of each object in the video had the best prediction capability amongst the static visual features \cite{han2014learning}.
Although including other semantically rich features in videos, such as emotions and actions are also effective \cite{sweeney2022overview}, high-level visual semantics, based on image captioning, was found to be the best predictor of both short-term and long-term memorability \cite{cohendet2018annotating, cohendet2019videomem}.
For example, SemanticMemNet \cite{newman2020multimodal} combines visual features and semantics through joint learning of caption generation to predict not only video memorability but also its rate of decay over time.
Although multimodal models, such as those combining visual and textual representations, perform best \cite{sweeney2022overview}, even a purely text-based model, based on sentence-level embeddings of short captions, performs quite well in predicting video memorability \cite{kleinlein2021topic}.
Longer and more descriptive texts were found to correspond to videos with high memorability \cite{guinaudeau2022textual}, indicating a correlation between the fidelity of semantic information and video memorability.
A more recently proposed video memorability prediction model, M3-S \cite{dumont2023modular}, attempts to emulate the steps of human memory formation, from encoding to understanding to consolidation, through four modules: raw perception, scene parsing, event understanding, and contextual similarity.
The high-level event understanding module, which handles action recognition, was found to be the most contributive to memorability prediction, again supporting the idea of memorability relying mostly on high-level semantics, with low-level features such as color and motion supplementing mostly by helping to differentiate between videos with similar semantics.
Results suggest that adding additional high-level features, such as emotion, may improve results even further, as action alone cannot represent complex semantics.

\subsubsection{Audio Memorability}
Like visual memory, auditory memory can be categorized into sensory, short-term, and long-term memory stores.
Since an auditory stimulus typically cannot be fully described by static features alone, the surrounding auditory context and acoustic patterns especially come into play for auditory memory \cite{winkler2005sensory}.
Auditory signals in sensory memory are transformed into more abstract, higher-level representations in short-term memory, which can then be attended to and even drive changes in perceptual sensitivity to further incoming stimuli \cite{zimmermann2016attending}.
Certain types of auditory stimuli appear to be remembered more than others, with spoken language recognition performing best, followed by sound objects, and lastly, musical excerpts, pointing to the significance of semantics in auditory memory as well \cite{cohen2009auditory}.

Memorability is also an intrinsic property of sound \cite{ramsay2018intrinsic}.
High-level conceptual features, especially causal uncertainty, visualizability, emotional valence, and familiarity, were found to be stronger predictors of sound memorability than low-level acoustic and salience features \cite{ramsay2018intrinsic}, similar to the way that high-level semantic features are the strongest predictors of visual memorability.
Essentially, memorable sounds tend to be familiar, emotional, easy to visualize, and come from apparent sources.
Although auditory memory is not as strong as visual memory \cite{cohen2009auditory, bigelow2014achilles}, audio-visual integration of semantically matching tracks has been found to enhance memory performance, even in the presence of temporal asynchrony \cite{meyerhoff2016semantic}.
Semantically matching environmental sounds, in particular, as opposed to verbalized words of the presented objects, enhance memory performance for the position of the object, likely due to triggering attentional mechanisms toward the location of the perceived sound source \cite{marian2021cross}.
Several video memorability prediction models explore the predictive power of audio features \cite{cohendet2018annotating, sweeney2020leveraging, kleinlein2020predicting, reboud2020predicting, sweeney2021predicting, sweeney2021influence, reboud2021exploring}, but most found little to no effect on overall video memorability.
Conditionally including audio features based on estimated audio gestalt memorability \cite{sweeney2021influence}, however, seems to be a promising approach.
High-level auditory features are used to determine the influence of the audio modality on overall memorability, and audio-augmented captions and audio spectrograms are included only if audio is predicted to be useful.
Recent findings confirm the dominance of visual long-term memory over auditory memory and provide evidence suggesting that auditory information is associated with visual information as soon as it is available, as opposed to being stored independently from or fully integrated with visual information \cite{meyerhoff2023long}, which may explain the indirect effect audio features have on video memorability.

Based on these findings, the following discusses our approaches to emulate some relevant aspects of memorability and their effects on KIS video retrieval tasks.

\section{Preprocessing Pipelines}
\label{sec:pipelines}
This section describes the design space and proposed pipelines for video preprocessing in the context of evaluating video retrieval systems in order to simulate some relevant aspects of human attention and memory.

\subsection{Design Space}
There are many ways to preprocess target videos to approximate the effects of human attention and memory.
In the context of video retrieval, it is important to consider what information is likely to be attended to and remembered and what information is useful for retrieval tasks.
Both the visual and audio tracks of a video can be filtered or transformed in different ways and at various levels of information preservation.
For this work, however, we limit ourselves to visual information and omit any auditory information to keep the design space more manageable.

As discussed in the previous section, semantic information is important for encoding events in visual long-term memory. Two possible ways to simulate memory effects may be to identify semantic information and either use it to filter the original video or transform the information into a new representation entirely through the synthesis of new content representing the same semantics differently. 

\subsubsection{Filtering}
The original target video can be filtered at different levels to mimic the effects of memory degradation.
We can take advantage of the findings by \citelong{cooper2019memories} to adjust color vibrancy to simulate the low-level degradation of scene representations in memory.
Global effects can be applied to the entire video, with resolution and color saturation adjustments applied evenly throughout, or applied in a vignette style to emulate the center bias effect of attention \cite{alvarez2005does}.
The applied filters could also be altered across the temporal domain to make forgettable segments appear more blurry and desaturated or even removed entirely through a frame filtering process.
For instance, memorability-based key frame selection and extraction have been successfully implemented in static \cite{fei2017memorable} and dynamic \cite{fei2018creating} video summarization methods.
Frame filtering has also been used to improve overall video memorability prediction \cite{constantin2022aimultimedialab}. Finally, one can adjust filters in the spatial dimensions by extracting meaningful foreground information through semantic segmentation or by emphasizing salient or memorable regions according to fixation or memorability maps, respectively.

\subsubsection{Synthesis}
Another idea is to abstract out the semantic content of the original videos to translate them into new ones.
Synthesizing new representations of the target videos can emulate both the loss of some pieces of information and the formation of new ones, taking into account false memory formation \cite{loftus1995formation}.
This could be done via manual methods such as sketching, reenacting, and animating based on the artists' memories of the target video.
The advantage of using manual methods is that the results can encapsulate real memory effects, as they are recreations based on the artist's actual memory of the original content.
Automatic approaches, on the other hand, have the advantage of being able to streamline the video synthesis pipeline.
This could be achieved by taking advantage of recent advances in generative models.
Budding image or video synthesis technologies can be used to create new content from the semantics of the original video.
Image generation is already impressive, with high-quality text-to-image technologies readily available to the public.
Text-to-video generation has evolved rapidly from GAN approaches to transformer-based frameworks and diffusion models in the past few years.
However, many of the state-of-the-art models such as Make-A-Video \cite{singer2022make}, Phenaki \cite{villegas2022phenaki}, NUWA-XL \cite{yin2023nuwa}, and OpenAI's Sora,\footnote{\url{https://openai.com/sora}} the latter three of which are able to produce longer, more complex, higher-quality videos, are not publicly available.

\subsection{Proposed Pipelines}
\begin{figure}[t]
\centering
\includegraphics[width=0.66\textwidth]{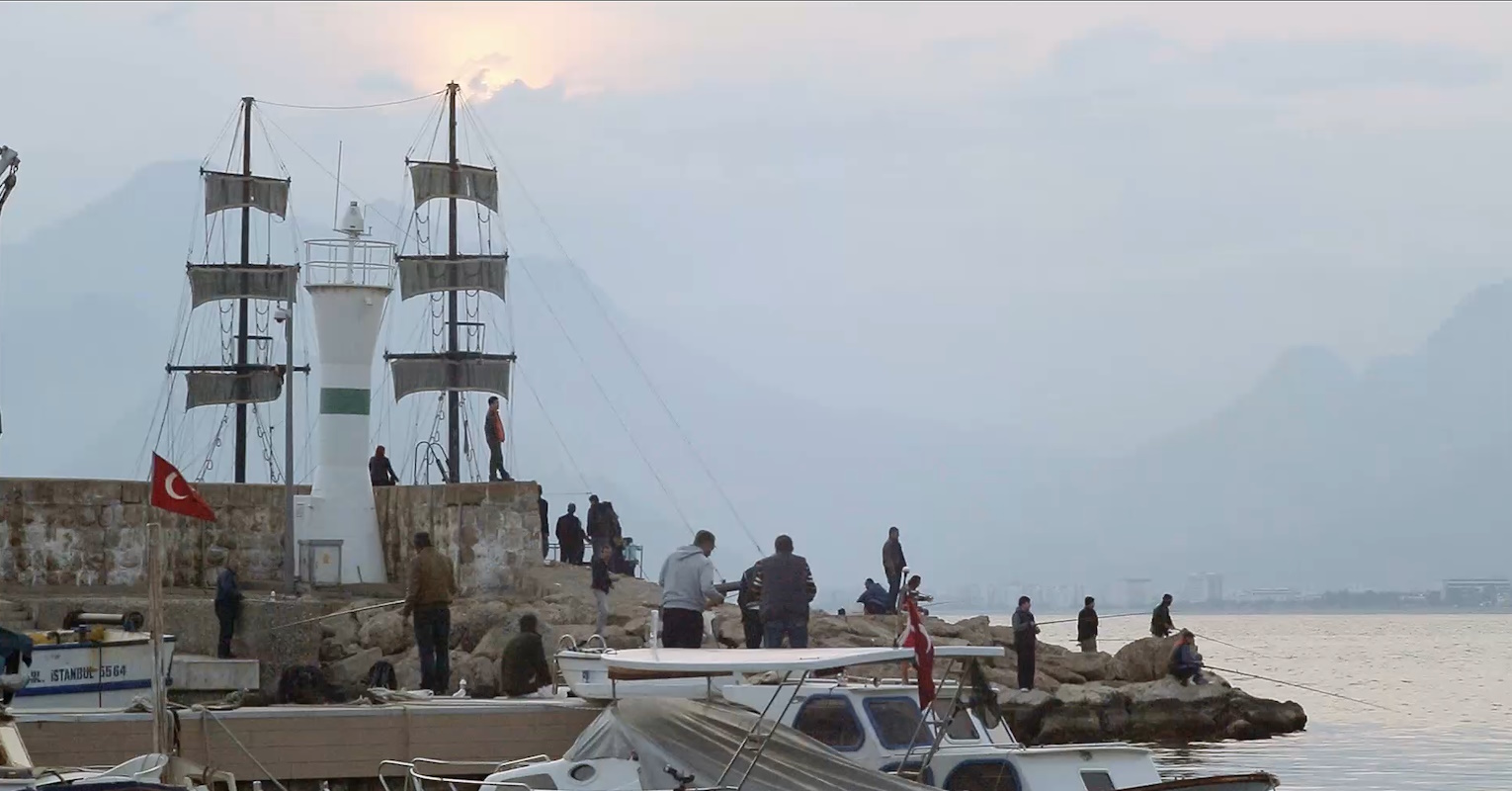}
\caption{Input Video}
\label{fig:original-video}
\end{figure}

We propose three filtering and three synthesis pipelines based on the design space discussed above.
We experiment with different levels of retained information to see how they affect the retrieval process.
Besides the global filtering pipeline and the text-to-video synthesis pipeline, we use video memorability as a high-level feature to filter or select frames.
Memorability estimations in these pipelines are obtained via AMNet \cite{fajtl2018amnet}, an attention-based image memorability estimation network that outputs memorability scores and attention maps, both of which can be used to filter video frames in different ways.
Figure \ref{fig:original-video} shows an unprocessed video frame, which we will use to showcase the effects of each pipeline.

\subsubsection{Video Filtering Pipelines}
The proposed filtering pipelines apply blur and desaturation in increasing levels of specificity, from global to frame to pixel-level granularity.

\paragraph{Global Filter (F1)}
A simple vignette effect is applied in the global filter, denoted as F1, so the outer regions are more blurred and desaturated than the video's center.

\begin{figure}[t]
    \centering
    \begin{subfigure}[b]{0.32\textwidth}
        \includegraphics[width=\textwidth]{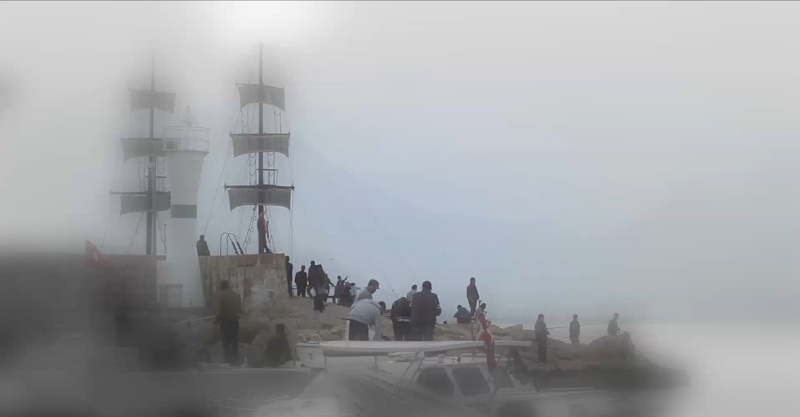}
        \caption{F1 Output}
    \label{fig:f1-output}
    \end{subfigure}
    \begin{subfigure}[b]{0.32\textwidth}
        \includegraphics[width=\textwidth]{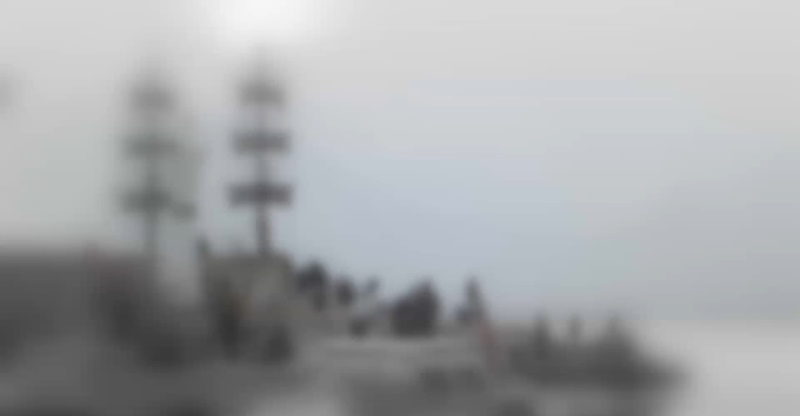}
        \caption{F2 Output}
        \label{fig:f2-output}
    \end{subfigure}
    \begin{subfigure}[b]{0.32\textwidth}
        \includegraphics[width=\textwidth]{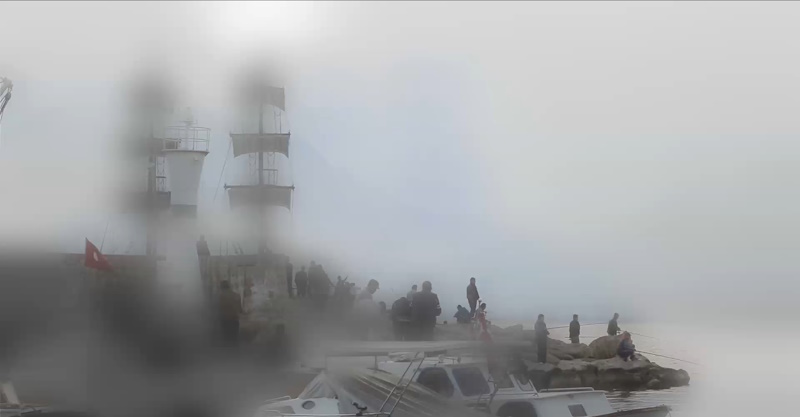}
        \caption{F3 Output}
        \label{fig:f3-output}
    \end{subfigure}
    
    \caption{Output of filtering pipelines}
\end{figure}

\paragraph{Frame Memorability Filter (F2)}
In addition to the global vignette filter, blur and desaturation are applied to the entire frame according to its estimated memorability score.
The degree of these effects may differ from frame to frame, but they are applied uniformly within each frame.
Inputting the original video into this pipeline,
gives us the output in Figure \ref{fig:f2-output}.

\paragraph{Spatial Memorability Filter (F3)}
In the spatial filtering pipeline, 
forgettable regions of the video, as estimated by generated memorability maps, are blurred and desaturated.
The values from the estimated memorability maps are first raised to the power of gamma, 0.8, and thresholded so that values below 0.4 are set to 0.
We then dilate the map with a dilation radius of 4 pixels and apply a Gaussian blur to smoothen the mask gradation.
A simple temporal smoothing is then applied with an alpha value of 0.6 to reduce jitter.
The amount of preserved information can thus differ within and across frames.
This is similar to the pipeline used in \cite{rossetto2021considering}, except with memorability masks instead of saliency masks.
An output of this pipeline is shown in Figure \ref{fig:f3-output}.


\subsubsection{Video Synthesis Pipelines}
The proposed synthesis pipelines below are described in order of increasing possibility for deviation from the original video.
Once an input video is segmented into its individual shots using the shot boundary detection framework TransNetV2 \cite{souvcek2020transnet}, each shot is input into SwinBERT, an end-to-end transformer-based model for video captioning \cite{lin2022swinbert}, unless video descriptions have been manually provided.
The generated or provided textual descriptions are then fed into a video generation model to produce a short video per shot, which is finally combined to form a sequence of shots that capture the semantics of the original video.
Two models are used for video generation: AnimateDiff \cite{guo2023animatediff}, a text-image-to-video generator, and Text2Video-Zero \cite{khachatryan2023text2video}, a zero-shot text-to-video diffusion model.
The former enables greater control over the output through an initial input image, whilst the latter only utilizes textual descriptions to synthesize videos.

\begin{figure}[t]
\centering

\begin{subfigure}[b]{0.32\textwidth}
    \includegraphics[width=\textwidth]{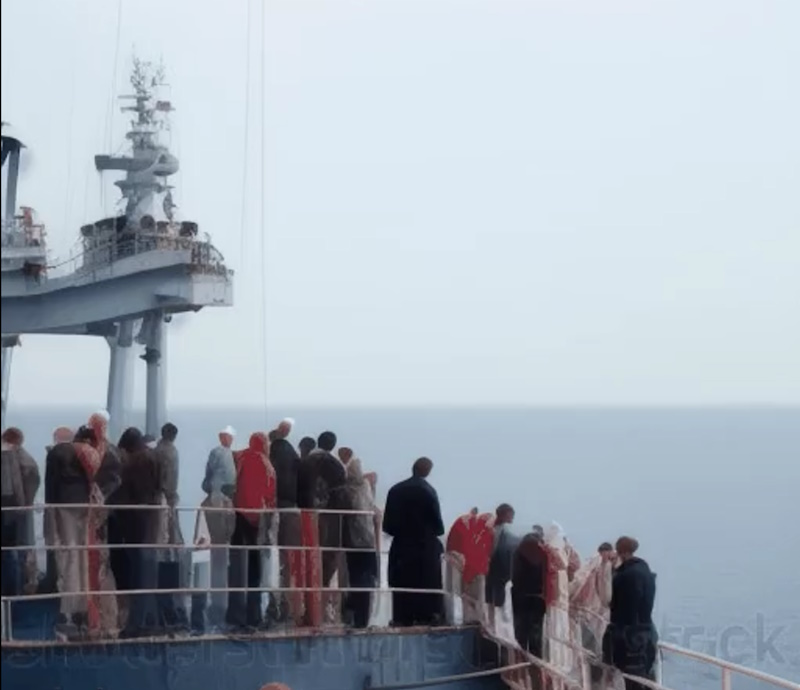}
    \caption{S1 Output}
    \label{fig:s1-output}
\end{subfigure}
\begin{subfigure}[b]{0.32\textwidth}
    \includegraphics[width=\textwidth]{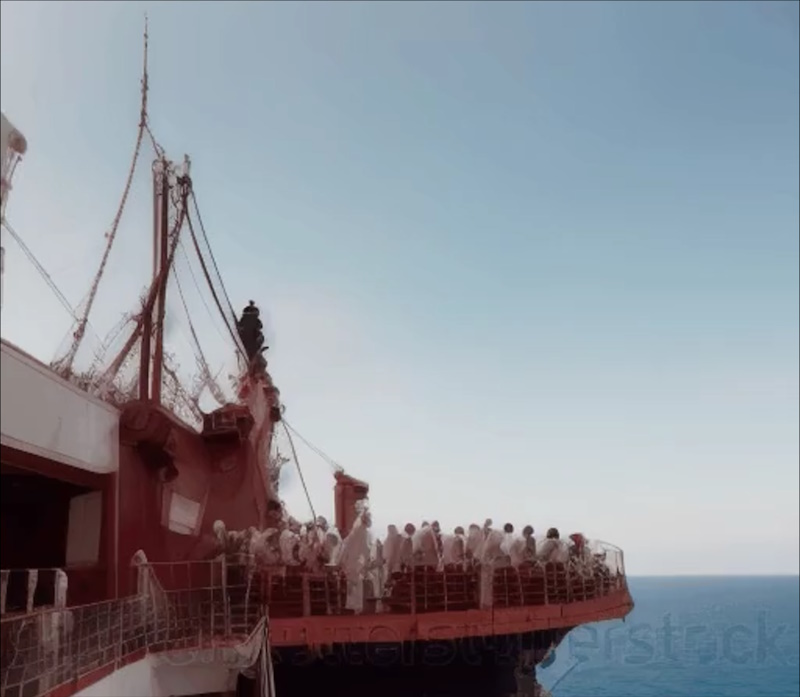}
    \caption{S2 Output}
    \label{fig:s2-output}
\end{subfigure}
\begin{subfigure}[b]{0.32\textwidth}
    \includegraphics[width=\textwidth]{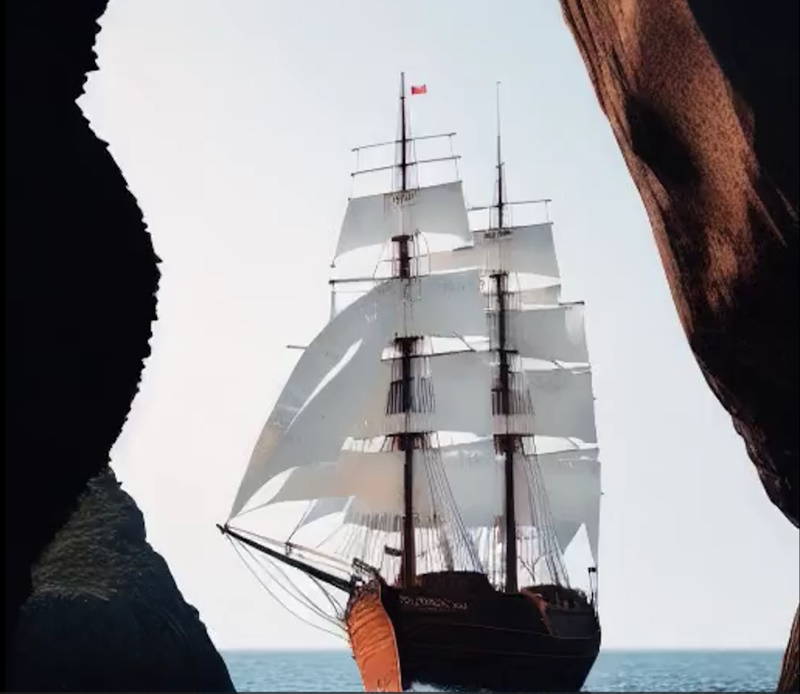}
    \caption{S3 Output}
    \label{fig:s3-output}
\end{subfigure}

\caption{Output of synthesis pipelines}
\end{figure}

\paragraph{Text \& Image to Video via Frame Selection (S1)}
The text and frame-to-video pipeline 
uses each shot's original frame and description to synthesize video segments.
For each shot, the first frame with a memorability score above a threshold, which we define as the average memorability score of the shot, is used as the starting frame.
Since the original frame is used in this scenario, the initial visualization should be closer to the original, while actions and resulting frames may deviate.
Figure \ref{fig:s1-output} shows an example output of this pipeline.


\paragraph{Text \& Image to Video via Image Generation (S2)}
In the text and synthesized image to video pipeline, 
the starting frame per shot is selected in the same way as above, but this time, novel images are generated, expanding the space of possible deviations from the original content.
For each selected frame, ControlNet \cite{zhang2023adding} is used to synthesize a new image based on its textual description and conditioned on its semantic map.
The synthesized starting image is then fed into a video generator along with the description of the shot to synthesize a video segment, such as the one shown in Figure \ref{fig:s2-output}.


\paragraph{Text to Video (S3)}
The final pipeline 
uses textual descriptions of shots directly to generate new versions of the shots.
This pipeline introduces the most possible noise and false information, as it relies on the text alone to capture original video semantics.
Unlike the previous two synthesis pipelines, there are no conditions restricting visual scene composition and details not mentioned in the description, which means that there is a greater likelihood for the output video to stray far from the original scene.
One output of this pipeline is shown in Figure \ref{fig:s3-output}.


\section{Pipeline and Query Selection}
\label{sec:preliminary}
This section describes the process of selecting the video targets and variations used in the final user experiment.
We began with an initial set of videos, using randomly selected tasks from the Video Browser Showdown (VBS) 2023 archive,\footnote{\url{https://github.com/lucaro/VBS-Archive}} and processed each with the pipelines described above.
The set was reduced through a qualitative selection process and then further refined based on the results of a preliminary survey gathered from a panel of video retrieval experts.
Throughout the rest of the paper, the shorthand abbreviation for each pipeline will be used to reduce verbosity.

\subsection{Initial Data Selection}
Ten videos were randomly selected from the Textual KIS tasks of the VBS 2023 archive.
All selected tasks use videos from the V3C video dataset~\cite{rossetto2019v3c}.
Videos were selected specifically from the Textual KIS task type because the textual descriptions provide a consistent source of human caption input to the synthesis pipelines.
These videos were clipped to the segments defined in the original task. They were each processed in nine different ways: through the three filtering pipelines, the three synthesis pipelines with automatic captioning, and the three synthesis pipelines with human captioning.

From the produced videos, qualitative selections were made between the human-captioned generated video and the machine-captioned one for each synthesis pipeline for each video.
Given two generated video clips per synthesis pipeline, we manually eliminated the less semantically accurate or more visually disturbing ones.
Some videos with automated captions, for instance, were wildly different in meaning from the original video due to inaccurate caption generation.
In some videos with manual captions, the descriptions were not granular enough to detail each shot, resulting in more generic videos.
A couple of example clips that were eliminated due to original semantics being ``lost in translation'' are shown in the Appendix, where Figure \ref{fig:video-06791-s} is synthesized from inputting Figure \ref{fig:video-06791} into the S2 pipeline, and Figure \ref{fig:video-13062-s} is synthesized from inputting Figure \ref{fig:video-13062} into the S3 pipeline, both using automatically generated captions.
Three out of the ten original queries produced comparatively inferior results for all synthesis pipelines, so they were removed from the candidate pool completely.

This process resulted in 49 video clips, with seven variations (original, three filtered, three synthesized) of seven different videos.

\subsection{Preliminary Study with Expert Users}
A preliminary study was conducted as a survey distributed to 28 researchers, most of whom are from the VBS community.
We asked them to evaluate the video clips in our selected set based on perceived visual or semantic similarity to help us select the most meaningful set for our experiment.

\paragraph{Filtered Videos}
For filtered videos, we wanted to ensure that each variation provided perceptibly different information from the original video and each other.
We showed participants the original video segment and the F1, F2, and F3 variations. We then asked them to \emph{`Please indicate which videos, if any, you would consider to be nearly visually identical (i.e., you gain equivalent information from the videos).`}
Choice options were provided in a matrix format, with subjects providing binary feedback.

Based on the results from this survey, the global vignette filter (F1) appeared to have the most similarities with other variations, especially to the original video and to the memorability-masked video (F3), as shown in Table \ref{tab:filtered-similarities}.
To reduce redundancy, we removed this variation in our final set. 

\begin{table}[]
\centering
\caption{Aggregated (binary) Similarity Votes Between Original and Filtered Videos by 28 Raters. The largest value is in bold, and the second largest underlined.}
\label{tab:filtered-similarities}
\begin{tabular}{|l||c|cc}
\hline 
Pipeline & Original     & \multicolumn{1}{c|}{F3}        & \multicolumn{1}{c|}{F2} \\ \hline \hline
F1       & \textbf{137} & \multicolumn{1}{c|}{{\ul 114}} & \multicolumn{1}{c|}{70} \\ \hline
F2       & 48           & \multicolumn{1}{c|}{104}       &                         \\ \cline{1-3}
F3       & 74           &                                &                         \\ \cline{1-2}
\end{tabular}
\end{table}

\paragraph{Synthesized Videos}
For the synthesized variations, the goal was to select variations that (1) are similar enough to the original video to make retrieval possible and (2) communicate different information from each other.
To address these, we had two types of questions for each video: (1) \emph{`Please indicate which video clips, if any, you would consider to be semantically similar to this video:'} followed by the original video clip, and (2) \emph{`Please indicate which of these video clips, if any, you would consider to be visually or semantically identical (i.e., there is no meaningful difference between them).'}
Survey participants could select multiple options for both question types among the S1, S2, and S3 video clips.
From the answers to these questions, no difference was found in the level of semantic similarity to the original video between S1, S2, and S3, each receiving 29.5\%, 26.9\%, and 33.6\% of the votes, respectively.
Rather, the videos that most people found to be more semantically close to the original depended heavily upon the video clip.

Between the synthesized variations, S1 (44\% of votes) and S2 (39.2\% of votes) were the most similar across the seven video clips.
This is unsurprising, considering S1 and S2 are generated from the same video synthesis model.

Since we want the synthesized variation to be similar enough to the original to provide meaningful information for retrieval, and there is no single synthesized variation that is consistently meaningful for all video clips, we selected only the variation that most people found to be most like the original for each video clip.
Five of the remaining seven videos were selected based on the number of votes for their most similar synthesized clip.

As a final result, this process resulted in two filtered versions (F2 and F3) and one synthesized version based on the best-rated pipeline per video.

\section{Experimental Setup}
\label{sec:setup}

In our main experiment, we want to determine the effects of the different ways of presenting the search target on the retrieval performance.
The experimental setup consists of a wrapper around the Distributed Retrieval Evaluation Server (DRES) \cite{rossetto2021system}, which provides KIS tasks and receives submissions, and the interactive video retrieval system, vitrivr \cite{rossetto2016vitrivr}, through which participants can search and send their submissions.
This section outlines the components of the platform and how they interact.

\subsection{Crowdsourcing Platform}
Prolific\footnote{https://www.prolific.co} is an online platform designed to enable large-scale data collection for academic research or machine-learning model training and evaluation.
We chose this platform over many others for its ease of use and thorough participant verification and bias mitigation processes.
When a participant starts the experiment, the participant's ID within Prolific is passed into our demographic survey on Qualtrics,\footnote{https://www.qualtrics.com} and then into our custom platform, where it is used to match a participant to their DRES credentials.
Once all tasks are complete, the participant is redirected back to Prolific with a completion code.

\begin{figure}[t]
\centering
\includegraphics[width=\textwidth]{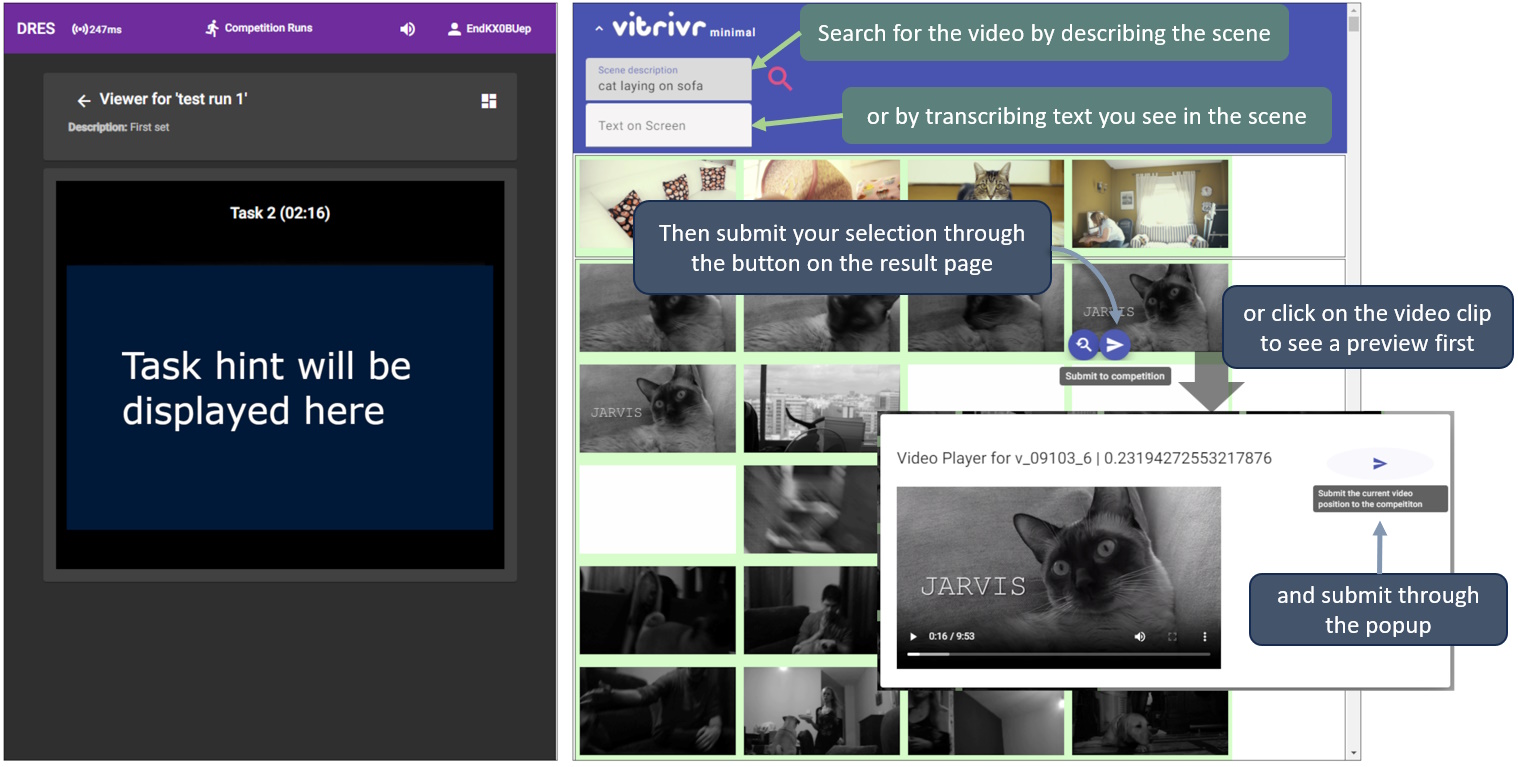}
\caption{Platform Screenshot}
\label{fig:platform-guide}
\end{figure}

\subsection{Experimental Platform}
After the demographic survey, there are three different stages that the participant can see on the frontend of the platform.
First is the starting screen, where instructions are provided.
Clicking the `Start' button on this screen triggers the backend to assign DRES credentials and a vitrivr instance to the participant.
Credentials are assigned randomly from a list of preregistered accounts.
If a participant has already been assigned to an account, their preexisting credentials will be reused such that they can pick up from where they left off.
These credentials are used to automatically log in to the DRES system and obtain the session token.
Since vitrivr may need to handle many queries at once, and a participant's performance in a task would rely heavily on the timeliness of vitrivr's response, it is necessary to ensure retrieval performance through load balancing.
Thus, the platform backend also handles assignment to vitrivr instances via a round-robin distribution of traffic.

Once the DRES credentials and vitrivr instance are assigned, the frontend renders two iframes, one with a viewer of the task from DRES and another containing vitrivr's minimal frontend.
The participant must search and submit their answers through the vitrivr iframe based on what they are shown from DRES until a correct submission is made or until the time is up, whichever comes first.
Figure \ref{fig:platform-guide} shows a screenshot of the main screen, annotated with instruction bubbles for the participant.
This is the image we display on the starting screen to familiarize participants with the task layout.

When all tasks are complete, the participant is shown the final screen, where they can click the `Finish' button to be directed back to Prolific and mark their completion.

\subsubsection{DRES}
As discussed in Section \ref{sec:related}, DRES facilitates the evaluation of interactive multimedia retrieval systems, including remote and asynchronous settings \cite{sauter2022asynchronous}.
This flexibility enables the crowdsourced exploration of different task presentations, which perfectly applies to our experimental setup.
From the administrative perspective, DRES allows us to configure an evaluation with a specific database, targets, task presentation, duration, and participants.
From the participant perspective, DRES displays the target, or a representation of the target, and receives participant submissions.
DRES provides an API that we use to make the participant's experience as smooth and seamless as possible.

\subsubsection{vitrivr}
The interactive video retrieval system we use in our platform is vitrivr \cite{rossetto2016vitrivr}.
This system offers a great variety of additive query modes. Still, we stick with their minimal frontend, vitrivr-ng-min~\cite{DBLP:conf/cbmi/SauterSWR23}, for the sake of simplicity and to avoid overwhelming potential novice users.
This frontend supports textual queries, namely text displayed on the screen and scene description.
Feature extraction must be conducted on the same video collection that was input to DRES before any video clip can be retrieved.
In addition to video search and retrieval, the vitrivr frontend supports making submissions to DRES directly through a button on the retrieved video segment.
The DRES session token obtained upon authenticating the participant is passed from our platform to vitrivr as soon as the vitrivr iframe is loaded to enable task submissions.

\subsection{Experimental Setup}
Using the final video set derived from the preliminary study, we set up five evaluations on DRES, one per experimental condition.
Each condition consists of five tasks, where each task is a unique video and variation combination, as summarized in Table \ref{table:tasks}.

\begin{table}[t]
\caption{Experimental Conditions}
\begin{tabular}{c*{5}{c}}
 & Task 1 (01140) & Task 2 (02024) & Task 3 (05722) & Task 4 (13872) & Task 5 (14700)   \\ 
\midrule
  Condition 1 & Original & F2 & F3 & S & Textual \\
  Condition 2 & F2 & F3 & S & Textual & Original \\
  Condition 3 & F3 & S & Textual & Original & F2 \\
  Condition 4 & S & Textual & Original & F2 & F3 \\
  Condition 5 & Textual & Original & F2 & F3 & S\\
\end{tabular}
\label{table:tasks}
\end{table}

Textual KIS tasks were included in the final evaluations to allow for comparison.
We used only the first textual hint from VBS to keep it consistent with the description we used as manual input to the relevant synthesis pipelines.
Table \ref{table:textual-hints} in the Appendix contains the descriptions of the videos in the final set.

We aimed to have 30-40 participants per condition and registered 50 members each to leave room to make up for incomplete or returned participation after DRES credentials assignment.
The platform design and study instructions were iterated from the feedback we received through test runs.
Having 500 distractor videos in the database and a duration of 3 minutes per task were set to balance the smaller pool of options with a shorter time limit.
These parameters were also confirmed to be sufficient in our test runs.
The participants we recruited for our study had a balanced distribution among males and females and were screened for English proficiency and high approval rates (80-100).

\section{Results}
\label{sec:results}
Excluding those who withdrew from the evaluation, did not give consent, or did not make any submission attempts, we gathered a total of 200 valid participants for this study.
For evaluations 1 through 5, corresponding to Table \ref{table:tasks}, we had 35, 43, 43, 36, and 43 participants, respectively.
Most participants indicated that they were familiar with using video retrieval systems.
The participants' technology interaction affinity scores, measured by the ATI scale \cite{Franke_Attig_Wessel_2018}, were also on the higher end, with more than 90\% of the participants having a score above 3.5 out of 6.
No correlation between ATI scores and video retrieval performance could be identified.

\begin{table}[t]
\centering
\caption{Total number of submissions per task type and submission result. We distinguish between submissions that are within the target segment, within 30 seconds or a minute of the target segment, within the target video, or outside the target video. The Tasks column shows the number of individual task instances. Since some participants did not complete all 5 tasks, the total number of task instances per type is below 200. Percentages are shown per task type (column-wise) and per total across task types and time intervals.}
\label{tab:results}
\resizebox{\textwidth}{!}{
\begin{tabular}{l|r|r|r|r|r||r|r|}
\cline{2-8}
                            & Correct      & Within 30s   & Within 1min & Within Video & Wrong        & Total        & Tasks \\ \hline
\multicolumn{1}{|l|}{Original} & 117 (26.8\%) & 103 (23.6\%) & 25 (5.7\%)  & 80 (18.3\%)  & 111 (25.5\%) & 436 (21.8\%) & 197 \\ \hline
\multicolumn{1}{|l|}{F2}       & 91 (26.4\%)  & 46 (13.3\%)  & 13 (3.8\%)  & 40 (11.6\%)  & 155 (44.9\%) & 345 (17.3\%) & 198 \\ \hline
\multicolumn{1}{|l|}{F3}       & 104 (22.7\%) & 96 (21.0\%)  & 41 (9.0\%)  & 108 (23.6\%) & 109 (23.8\%) & 458 (23.0\%) & 199 \\ \hline
\multicolumn{1}{|l|}{S}     & 6 (2.1\%)    & 37 (13.0\%)  & 25 (8.7\%)  & 54 (18.8\%)  & 165 (57.5\%) & 287 (14.4\%) & 198   \\ \hline
\multicolumn{1}{|l|}{Textual}  & 33 (7.0\%)   & 109 (23.1\%) & 50 (10.6\%) & 81 (17.2\%)  & 199 (42.2\%) & 472 (23.6\%) & 196 \\ \hline \hline
\multicolumn{1}{|l|}{Total} & 351 (17.6\%) & 391 (19.7\%) & 154 (7.7\%) & 363 (18.2\%) & 739 (37\%)   & 1998         & 988   \\ \hline
\end{tabular}
}
\end{table}

\paragraph{Answer correctness}
Table~\ref{tab:results} shows the submissions that are correctly within the target segment, miss the target segment by at most 30 seconds, between 30 seconds and one minute, and more than one minute, as well as the submissions that missed the target video entirely. It also shows the total number of submissions per task type as well as the total instances per task type.
A detailed breakdown per task variant is shown in Table~\ref{tab:submissions-by-task} in the appendix.
The fractional numbers show the ratio between correctness types per task type as well as the proportion of submissions per task type with respect to the total number of submissions.
Since a few participants did not complete the entire experiment, we were only able to obtain slightly below the targeted 200 instances per task type.
It is important to note that while a participant could make an arbitrary number of incorrect submissions per task, the task would end after a correct submission.

We can see that the unmodified video target was found most often, followed by the two filtering pipelines.
The lowest number of correct submissions, in both absolute and relative terms, as well as the lowest number of total submissions, can be seen for the tasks using a synthetic video target.
The tasks with textual targets, while also having a comparatively small number of correct submissions, have a much larger number of near-misses when compared to the synthetic targets.
The general success rate for solving any task, i.e., the number of correct submissions divided by the total number of tasks per type, ranges from 59.4\% for unmodified targets to 3\% for synthetic targets.

\paragraph{Time for task completion}
When considering only the correct submissions, the distribution of the times taken from the start of a task until the participants could find the correct segment can also provide some insights into the difficulties of the task types.
The mean time in seconds within one standard deviation for the different task types is as follows: $97.6 \pm 38.7$s for the Original targets, $105.3 \pm 43.3$s for F2, $96.3 \pm 38.9$s for F3, $131.1 \pm 26.1$s for S, and $118.3 \pm 34.4$s for the Textual tasks.
We can see no substantial difference in the times for the Original and F3 tasks, with F2 tasks taking only slightly longer to solve.
Tasks with synthetic video targets needed by far the longest time to be solved, with, on average, over 2 out of the available 3 minutes per task.

\paragraph{Query terms}
Looking at the search terms corresponding to wrong video submissions, some common patterns include terms that are not related to the video hint itself (e.g., `Second set', `Competition Runs'), descriptions that are too general (e.g., `music', `wedding', `race'), and terms that focused on aspects of the provided hint that are not relevant to the actual target (e.g., `kayak blurry', `AI generated climber').

\begin{figure}[t]
  \centering
  \begin{subfigure}[b]{0.33\textwidth}
    \includegraphics[width=\textwidth]{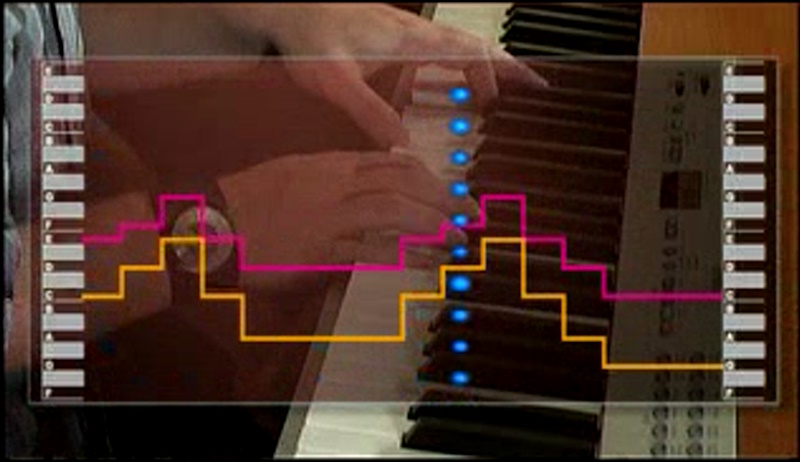}
    \subcaption{Target}
    \label{fig:video-02024}
  \end{subfigure}
  \begin{subfigure}[b]{0.22\textwidth}
    \includegraphics[width=\textwidth]{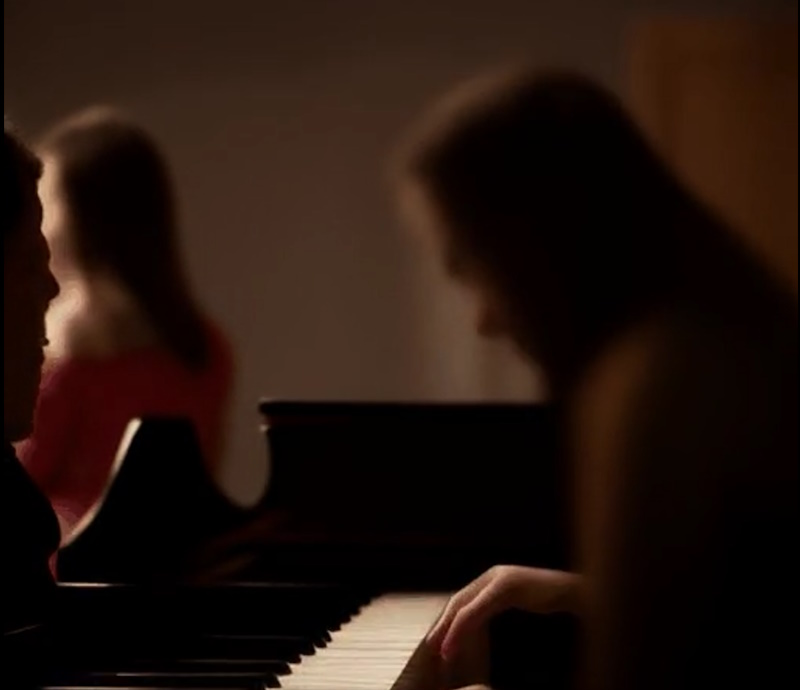}
    \caption{Hint}
    \label{fig:synth-02024}
  \end{subfigure}
  \begin{subfigure}[b]{0.33\textwidth}
    \includegraphics[width=\textwidth]{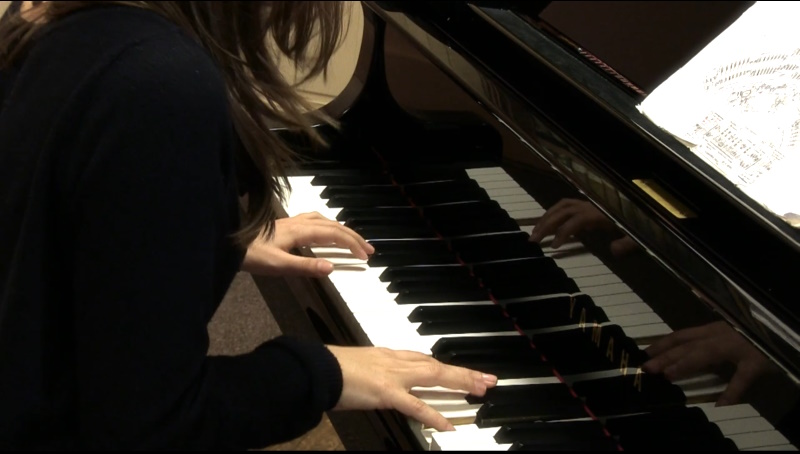}
    \subcaption{Common Submission}
    \label{fig:14607}
  \end{subfigure}
  \caption{Most Common Wrong Video Submission for Video 02024, given a Synthesized Video Hint}
\end{figure}

\paragraph{Most common failure}
The most commonly submitted incorrect video amongst all task types and videos is Video 14607 given a synthesized variation of the target Video 02024.
This is, in fact, the only synthesized video task that received no correct submissions.
Taking a closer look at this, we can see that Video 14607 appears to be visually closer to the synthesized task hint than the synthesized video is to the target video.
Figure \ref{fig:video-02024} shows a frame of the original target video.
The synthesized task hint is displayed in Figure \ref{fig:synth-02024}, and a frame of the commonly submitted Video 14607 is shown in Figure \ref{fig:14607}.
This says more about the quality of the synthesized video than the ability of participants to extract relevant semantic information from the hint provided.
Limitations in different stages in the synthesis process may make identifying and representing more complex qualities, such as transparent graphical overlays, difficult.

\section{Discussion}
\label{sec:discussion}
The findings emerging from the user experiment offer valuable insights into the effect of different target representations on the overall retrieval performance.
In this section, we examine the implications of these findings, identify the limitations inherent in this study, and explore potential avenues for future work in the domain of interactive media retrieval evaluations.

\subsection{Interpretation}
The filtered video variations in this study exhibit a level of comparability to the original Visual KIS task. In contrast, the synthesized variations currently demonstrate a greater degree of comparability to the Textual KIS task.
Results from the user experiment provide us with an initial insight into the range and effects of different task presentation methods and what this could mean in relation to our cognitive processes.

From the results, we can see that filtering a target video to blur and desaturate less memorable frames (F2) or regions (F3) can produce similar video retrieval results to showing the target as a hint itself.
In general, F2 variations tend to appear blurrier than F3 variations due to the entire frame being blurred at the same level, while F3 variations appear to spotlight more memorable regions.
This could potentially be related to why a greater proportion of wrong submissions in the F2 task type are not even within the correct video, as more participants could be searching with terms related to the blurriness of the F2 video. 

Although the proportions of incorrect submission types differ between F2 and F3 task types, it is noteworthy that the F2 task received a good proportion of correct submissions, comparable to the unfiltered task hint, and slightly surpassing F3.
Upon a closer examination of individual task results, participants generally outperformed on F3 tasks compared to F2 tasks.
However, there are a couple of exceptions, namely video 14700, where F2 garnered more correct submissions but fewer near-correct submissions than F3, and video 02024, where F2 received a significantly higher proportion of correct submissions (48.8\%) compared to F3 (14.4\%).
In these instances, perhaps the overall blur of F2 filtering reduced the prominence of distracting features within the video, or the effects of F3 filtering were more pronounced in specific areas of the video.
The latter would suggest that regions estimated to be forgettable were, in fact, containing elements people used in their search queries.
Generally, \emph{the higher prevalence of incorrect submissions in filtered tasks compared to the original unfiltered task type indicates some information loss due to filtering that affects video search}.

The synthesized task type resembles the textual task because three of the synthesized video hints directly mirror the textual hints of their respective videos, and the participants face greater difficulty in pinpointing the correct video segments in both scenarios.
In both cases, the hint is more abstract, relying heavily on the participant's interpretation of the information provided.
Although the synthesized and textual task hints have similar submission accuracies in general, submissions for synthesized task types are slightly worse overall, suggesting that the \emph{artificially generated visualizations may either be distracting, causing participants to focus more on its visual aspects than the semantic ones, or portraying new or inaccurate semantic information, causing participants to misinterpret the idea of the target}. 

Two cases comparing synthesized and textual tasks stand out in particular.
One is video 02024, where the textual task had some correct submissions and many incorrect submissions that were still within the target video.
In contrast, the synthesized version had no correct submissions, and all incorrect submissions were from outside the target video.
This disparity likely stems from a significant missing element in the synthesized video -- the absence of the colorful graphical overlay.
Complex videos, such as those with transparent layers, may lose some meaning when processed through the synthesis pipelines due to the limited capabilities of the models themselves.
Another interesting case is video 05722, where the textual task yielded no correct submissions or even submissions within the target video.
The synthesized video, on the other hand, had a small portion of submissions that were correct within a minute of the target segment or at least within the same video.
This could be a case where the synthesized hint provides something meaningful and useful beyond text alone. 

While \emph{both textual and synthesized task representations extract and portray the semantics of a video, the synthesized variations take it a step further by reinterpreting the extracted semantics}.
The synthesis pipelines' premise was that high-level semantics, the aspect most crucial to memorability, are effectively captured by textual descriptions.
Generating a visualization based on the text yields a common interpretation of the semantics for all participants.
In contrast, textual KIS tasks rely more on individual interpretation and can be influenced by personal experience, context, and language proficiency.
\emph{Through video synthesis mechanisms, we can provide a common visualization and fill in visual details in potentially inaccurate ways, bridging gaps with `fake memories'} that are consistently simulated across participants.
One challenge, however, is determining the degree of deviation from the original source that replicates real memory effects.
Various methods of synthesizing videos were explored, including using manual or automatic captions, frame-based video generation, frame-to-image synthesis with semantic mapping, and text-to-video synthesis.
No one method consistently outperformed the others, varying instead from video to video.
This variability could be attributed to this specific set of videos or may stem from technological limitations.
Nonetheless, a more comprehensive study on video synthesis would be needed to determine the most appropriate method for this use case.

In summary, we found that different ways of presenting the target of a retrieval task have a clear effect on overall retrieval performance.
Applying filtering pipelines that aim to simulate effects similar to those caused by attention and memory shows a slight decrease in retrieval performance, indicating that such effects should be considered in interactive retrieval evaluations.
Using filtered versions of the original video did, however, still lead to a higher rate of solved tasks compared to the use of a textual description.
This suggests that finding the target based on even partial visual information is easier than having to imagine the target based on a textual summary alone.
Using a synthetic target appears to make the task even more difficult, even though it alleviates the differences caused by different users imagining the target in different ways.
This type of target representation can be interpreted as a way of causing a consistent way of misremembering a target, which deserves further study.

This has implications for the task design of evaluations as it questions the external validity of some of the designs. 
Someone aiming to find a video from memory appears to be a vastly different task than being told to find some video containing some scene based on a textual description or even the task of finding something similar to some given artifact.
Hence, engaging in use-case studies of the actual usage of multi-media retrieval methods would be valuable in informing the next generation of benchmark design.

\subsection{Limitations}
The constraints in this work mainly stem from technology or scope.
As we relied on open-source materials for constructing different experimental pipelines, we were limited by the technologies that are currently openly available.
The second type of limitation is more intentional.
As there are many possible paths we can take in an exploratory study, we needed to select which areas to apply our focus, inevitably leaving out other aspects for future work.

As discussed previously, this study explored artificially generated video clips, among other methods, to convey hints in KIS tasks.
At this stage, however, it is crucial to acknowledge that the openly available video synthesis technologies exhibit certain limitations.
The generated clips often suffered from distortions and instability, with objects seemingly morphing rather than moving smoothly over time.
Such anomalies can potentially confuse viewers or divert their attention toward the unintended visual effects rather than the intended semantic content.
Given the rapid advancements in the field of video synthesis and the growing body of research focused on improving these techniques, it may be worthwhile to revisit this approach in the future, exploring different, more advanced models that can mitigate these issues. 

Another technical constraint arises from the memorability prediction method.
The availability of pretrained video memorability models, which also enable the extraction of spatial mappings, remains limited.
Predicting video memorability with precision continues to be a considerable challenge.
While this study incorporated an image memorability estimation model into the pipelines, there is potential merit in investigating the use of a video-specific model to enhance the accuracy and stability of predicting memorable regions within videos.

Constraints on what conclusions can be drawn also arise simply from the limited scope of this study.
The pipelines used to preprocess KIS task hints are derived from heuristics and models based on research on human perception and memory but are not independently validated to be accurate representations of such effects.
While we can see that each pipeline produces a unique effect on video retrieval success, it remains uncertain which, if any, most faithfully emulates real memory effects.
Additionally, we focused on visual and automated approaches, but numerous other methods can be explored.

\subsection{Future Work}
This work can be expanded upon in several different ways.
For instance, one could go in the direction of validating whether or not any of the presented pipelines effectively capture real perception and memory effects.
It could also be worth continuing to explore other task presentation methods and evaluation protocols.

A logical progression in this research would involve an experimental examination of the influences of perception and memory on video retrieval.
In such an experiment, participants would be presented with the actual target segment and then asked to retrieve it after a certain amount of time has elapsed.
The resulting findings could then be compared to the outcomes of this study, shedding light on which pipeline most faithfully captures the effects of perception and memory.
There are many well-known studies on human memory and even video memorability, but they are conducted by re-showing videos or using other recall prompts \cite{shekhar2017show, cohendet2018annotating, cohendet2019videomem}.
It would be interesting to conduct analogous investigations within the context of interactive retrieval, leveraging an evaluation server and content-based media retrieval system such as DRES and vitrivr.

Another avenue for future research is to explore alternative methods of task presentation.
While this paper concentrated on automated approaches to representation synthesis, numerous manual methods exist that could be considered, such as engaging artists to sketch, reenact, or animate a target video from memory.
These representations, grounded in an individual's human interpretation and memory of a video, might yield more informative hints for video retrieval and offer a closer approximation to real memory effects.

Diverging from the visual domain, another promising area for exploration could revolve around audio, which was intentionally excluded from this study to maintain a focused scope.
Although audio is not as predominant in memory as visuals, its impact on the video retrieval process could be investigated. Possibilities include applying a basic band-pass filter, experimenting with noise addition, and delving into the memorability of sound.
Exploration of the latter could involve identifying and diminishing less memorable audio features while amplifying the more memorable ones, such as speech \cite{cohen2009auditory}.
In general, one might take different high-level aspects of sound memorability into consideration, such as sound source clarity, emotional valence, and familiarity \cite{ramsay2018intrinsic}.
Examining visual-audio interactions \cite{meyerhoff2023long}, such as visualized sound cues and sources, adds another layer of complexity to this investigation.

\section{Conclusions}
\label{sec:conclusion}
This paper presented an exploratory investigation into the integration of perceptual and memory effects within interactive video retrieval tasks.
Concepts from the literature were applied through the design and implementation of video processing pipelines, focusing on visuals and automated approaches.
Six different pipelines emerged from this process: three of which filter the original target segment in various ways, and three synthesize new videos from the original input using generative models.
A focus was placed on memory effects, as they also implicitly encapsulate influences of perception and attention.

We then created a custom experimental platform and used it to conduct a crowdsourced known-item search evaluation that employed specific video variations, which were generated through the pipelines and selected through a preliminary evaluation.
These various task representations have varying effects on the video retrieval process.
Still, in general, people can find the target video despite obscuring, removing, or even altering large parts of the video shown as a hint.
We found that the way in which task targets are presented impacts overall retrieval performance.
While filters aiming to simulate attention and memory effects appear to make the task slightly more difficult, the differences are small compared to using textual or synthetic visual task targets.
The findings might have consequences for the design of future benchmarks, as they show the importance of considering additional aspects of real-world scenarios after which the benchmark is modeled.

To the best of our knowledge, the experiment presented in this paper was, by a substantial margin, the largest interactive video retrieval evaluation in terms of the number of participants conducted so far.

The extent to which these representations accurately capture human memory effects remains a subject for future investigation.
Nevertheless, the results showcased in this study are highly promising.
They underscore the significant influence of target representation on retrieval success and demonstrate the malleability of this influence through different approaches.
It is highly plausible that real perception and memory effects would introduce nuances in retrieval success rates that are not captured by the current practices of evaluating known-item search tasks in media retrieval evaluations.
It would be interesting to see a more realistic scenario being emulated in large-scale evaluations and how this might influence evaluation outcomes.

\begin{acks}
The authors would like to thank all the participants in the preliminary study.
This work was partially supported by the Swiss National Science Foundation through project MediaGraph (contract no.\ 202125).
\end{acks}

\bibliographystyle{ACM-Reference-Format}
\bibliography{bibliography}

\appendix

\pagebreak

\section{Online Appendix}

\subsection{Additional Figures and Tables}

\begin{table}[h]
\caption{Textual Hints}
\begin{tabular}{p{0.10\linewidth} | p{0.85\linewidth}}
 ID & Description  \\ 
\midrule
  01140 & Start of an indoor bike race with 6 riders. A motorbike with a camera crosses the start line just after the starting shot. \\
  02024 & Singing instruction video, showing two singers and a keyboarder, with an overlaid graphical visualization. \\
  05722 & Shot of a wedding party panning from left to right, the party is grouped around bride and groom, then a shot of bride and groom walking and guests following them. \\
  13872 & Kids in kayaks on a river, throwing paddles through three coloured hoops placed over the water. \\
  14700 & View down the surface of a boulder, with a forest in the background. A bearded man in a cyan shirt climbing up the boulder. \\
\end{tabular}
\label{table:textual-hints}
\end{table}

\begin{table}[]
\centering
\caption{Submission Accuracies Per Task}
\label{tab:submissions-by-task}
\begin{tabular}{|l|l||r|r|r|r|r|}
\hline
Video &
  Pipeline &
  \multicolumn{1}{l|}{Correct} &
  \multicolumn{1}{l|}{Within 30s} &
  \multicolumn{1}{l|}{Within 1m} &
  \multicolumn{1}{l|}{Within Video} &
  \multicolumn{1}{l|}{Wrong} \\ \hline \hline
\multirow{5}{*}{01140} & Original & 83.3\% & 0\%    & 4.2\%  & 8.3\%  & 4.2\%  \\ \cline{2-7} 
                       & F2       & 32.4\% & 1.5\%  & 0\%    & 0\%    & 66.2\% \\ \cline{2-7} 
                       & F3       & 39.2\% & 7.8\%  & 0\%    & 0\%    & 52.9\% \\ \cline{2-7} 
                       & S        & 6.5\%  & 0\%    & 0\%    & 16.1\% & 77.4\% \\ \cline{2-7} 
                       & Text     & 6.5\%  & 15.2\% & 0\%    & 2.2\%  & 76.1\% \\ \hline \hline
\multirow{5}{*}{02024} & Original & 28.4\% & 7.8\%  & 4.9\%  & 46.1\% & 12.7\% \\ \cline{2-7} 
                       & F2       & 48.8\% & 7.3\%  & 0\%    & 31.7\% & 12.2\% \\ \cline{2-7} 
                       & F3       & 14.4\% & 0.8\%  & 2.4\%  & 62.4\% & 20.0\% \\ \cline{2-7} 
                       & S        & 0\%    & 0\%    & 0\%    & 0\%    & 100\%  \\ \cline{2-7} 
                       & Text     & 20.0\% & 1.7\%  & 0\%    & 53.3\% & 25.0\% \\ \hline \hline
\multirow{5}{*}{05722} & Original & 19.6\% & 7.2\%  & 0\%    & 12.4\% & 60.8\% \\ \cline{2-7} 
                       & F2       & 22.9\% & 13.5\% & 8.3\%  & 15.6\% & 39.6\% \\ \cline{2-7} 
                       & F3       & 55.0\% & 10.0\% & 5.0\%  & 17.5\% & 12.5\% \\ \cline{2-7} 
                       & S        & 4.5\%  & 0\%    & 2.3\%  & 4.5\%  & 88.6\% \\ \cline{2-7} 
                       & Text     & 0\%    & 0\%    & 0\%    & 0\%    & 100\%  \\ \hline \hline
\multirow{5}{*}{13872} & Original & 31.0\% & 56.0\% & 6.0\%  & 2.4\%  & 4.8\%  \\ \cline{2-7} 
                       & F2       & 19.8\% & 22.1\% & 4.7\%  & 3.5\%  & 50.0\% \\ \cline{2-7} 
                       & F3       & 36.8\% & 35.5\% & 5.3\%  & 7.9\%  & 14.5\% \\ \cline{2-7} 
                       & S        & 4.2\%  & 41.7\% & 25.0\% & 4.2\%  & 25.0\% \\ \cline{2-7} 
                       & Text     & 8.7\%  & 35.6\% & 24.0\% & 4.8\%  & 26.9\% \\ \hline \hline
\multirow{5}{*}{14700} & Original & 17.8\% & 31.8\% & 10.9\% & 13.2\% & 26.4\% \\ \cline{2-7} 
                       & F2       & 18.5\% & 18.5\% & 1.9\%  & 16.7\% & 44.4\% \\ \cline{2-7} 
                       & F3       & 9.6\%  & 36.1\% & 19.3\% & 10.2\% & 24.7\% \\ \cline{2-7} 
                       & S        & 0.9\%  & 25.0\% & 16.7\% & 42.6\% & 14.8\% \\ \cline{2-7} 
                       & Text     & 5.6\%  & 4.0.\% & 15.6\% & 26.9\% & 11.9\% \\ \hline
\end{tabular}
\end{table}

\begin{figure}[h]
  \centering
  \begin{subfigure}[b]{0.30\textwidth}
    \includegraphics[width=\textwidth]{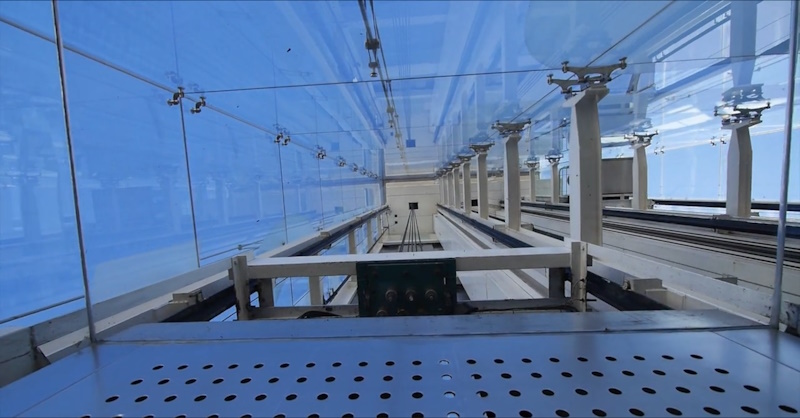}
    \subcaption{Frame from video 06791}
    \label{fig:video-06791}
  \end{subfigure}
  \hfill
  \begin{subfigure}[b]{0.18\textwidth}
    \includegraphics[width=\textwidth]{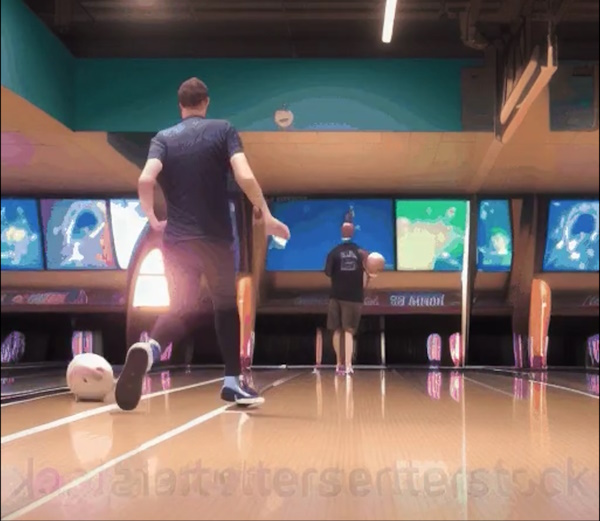}
    \subcaption{S2 Variation}
    \label{fig:video-06791-s}
  \end{subfigure}
  \hfill
  \begin{subfigure}[b]{0.30\textwidth}
    \includegraphics[width=\textwidth]{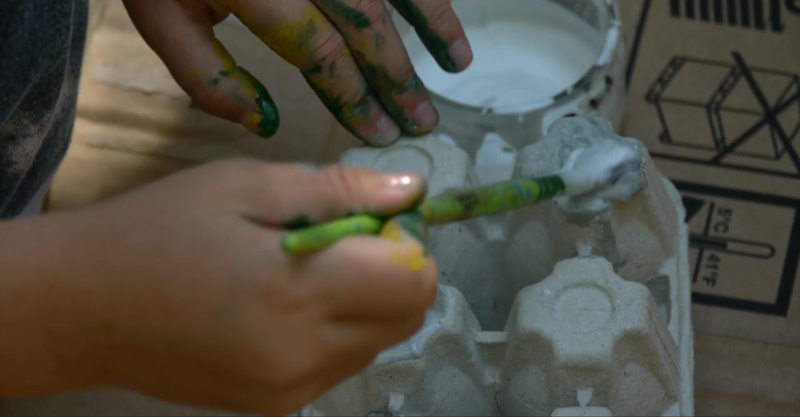}
    \subcaption{Frame from video 13062}
    \label{fig:video-13062}
  \end{subfigure}
  \hfill
  \begin{subfigure}[b]{0.18\textwidth}
    \includegraphics[width=\textwidth]{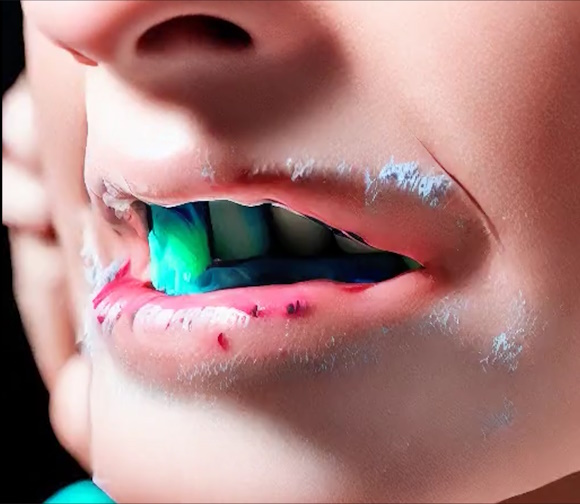}
    \caption{S3 Variation}
    \label{fig:video-13062-s}
  \end{subfigure}
  \caption{Examples of eliminated synthesis outputs}
\end{figure}



\end{document}